\documentclass{PoS}

\usepackage{graphicx}
\usepackage{txfonts}

\title{On the physical meaning of the 2.1 keV absorption feature in 4U 1538$-$52}

\ShortTitle{On the physical meaning of the 2.1 keV absorption feature in 4U 1538$-$52}

\author{\speaker{Jos\'e J.~Rodes Roca}\thanks{This work was supported by the MICINN project number AYA2010-15431.}, Jos\'e Miguel~Torrej\'on, Silvia~Mart\'{\i}nez-N\'u\~nez and Guillermo Bernab\'eu\\
        Universitat d'Alacant, Spain\\
        E-mail: \email{rodes@dfists.ua.es}}

\abstract{The improvement of the capabilities of nowadays X-ray observatories, like \emph{Chandra} or
   \emph{XMM-Newton}, offers the possibility to detect both absorption and
   emission lines and to study the nature of the matter surrounding the neutron star in X-ray binaries and
   the phenomena that produce these lines.
The aim of this work is to discuss the different physical scenarios in order to explain
the meaning of the significant absorption feature present in the
   X-ray spectrum of 4U 1538$-$52. Using the last available calibrations, we discard the possibility that this feature is due to calibration,
   gain effects or be produced by the X-ray background or a dust region. Giving the energy resolution of the \emph{XMM-Newton} telescope
we could not establish if the line is formed in the atmosphere of the neutron star
or by the dispersion of the stellar wind of the optical counterpart.}

\FullConference{An INTEGRAL view of the high-energy sky (the first 10 years)"
9th INTEGRAL Workshop and celebration of the 10th anniversary of the launch,\\
		October 15-19, 2012\\
		Bibliotheque Nationale de France, Paris, France}

\begin{document}

\section{Introduction}

Thanks to the high-sensitivity and high-resolution instruments on board
    current X-ray observatories, like \emph{XMM-Newton}, \emph{Chandra} or
    \emph{Suzaku}, it is possible to observe both emission and absorption
    lines in many X-ray sources: High Mass X-ray Binaries, Low
    Mass X-ray Binaries or Active Galactic Nuclei. 

    The detection and identification of spectral features are a tool to infer,
    for example, the chemical composition of the neutron star (NS) atmosphere,
    densities of the illuminated gas in the stellar wind or to estimate the
    magnetic field of the NS.

    Absorption features at 0.7 keV, 1.4 keV and 2.1 keV have been detected in
    the isolated NS 1E 1207.4$-$5209 and several different interpretations were
    proposed (\cite{sanwal02,mereghetti02,bignami,dl04,moha06,xu11}). In fact,
    depending on the spectral analysis of the \emph{XMM-Newton} data,
    the feature at 2.1 keV is interpreted either as electron cyclotron absorption line
    (\cite{bignami,dl04, suzaku}) or as of instrumental origin (\cite{mori05}).

    In this work, we discuss the implications of the different interpretations
    of the 2.1 keV feature in 4U 1538$-$52.

\section{Observation}

The \emph{XMM-Newton} observation of 4U 1538$-$52 started on
August 14 and finished on August 15, 2003. The details of the
data reduction can be found in \cite{RR11}.
We found strong residual structure around 2.1 keV
and checked the instrumental origin with the \emph{EPIC/pn} calibration
team. Different possible formation mechanisms are discussed.

\section{Spectral analysis}

In order to identify new features, we have carried out a similar process
        to the one described in \cite{RR11}. The first explanation for the
        presence of such absorption feature would be a bad correction of the 
        well known decrease in the effective area near 2 keV due to the Au M edge.
        Therefore, we have conducted a systematic study to discard
        completely the instrumental origin.

\begin{figure*}
\centering
\includegraphics[width=.4\textwidth]{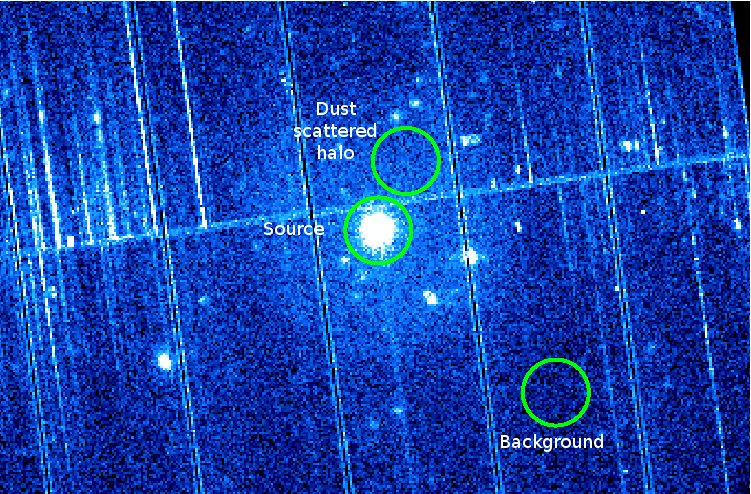}
\includegraphics[width=.4\textwidth]{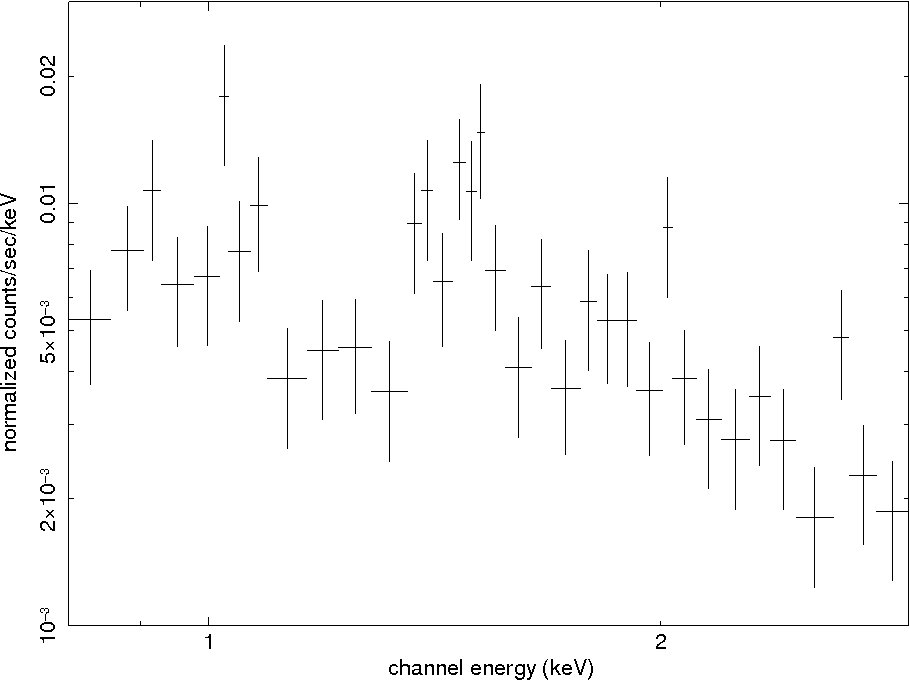}
\includegraphics[width=.4\textwidth]{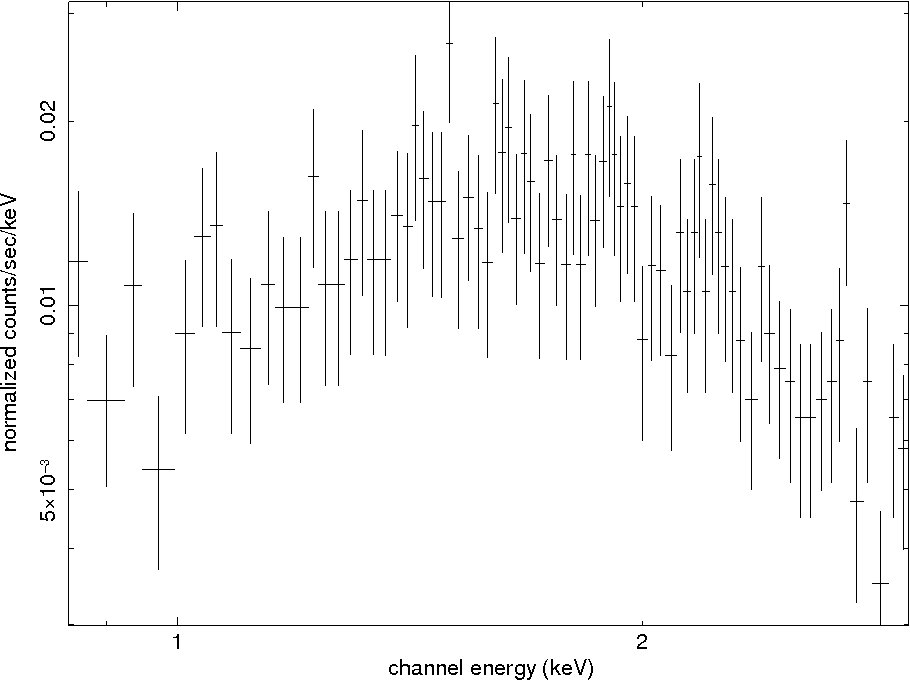}
\includegraphics[width=.4\textwidth]{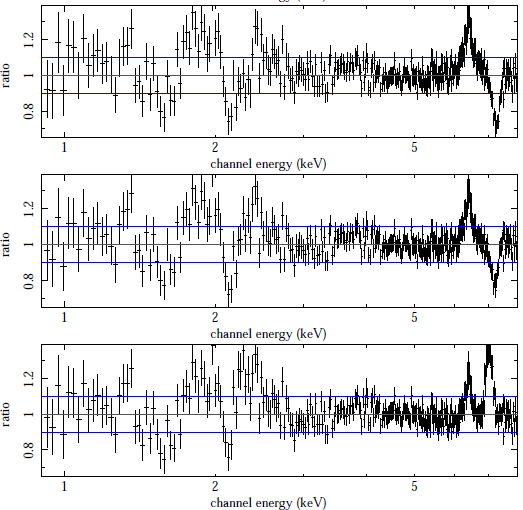}
\caption{The circles show the extraction region (\emph{left top}) from
the background spectrum (\emph{right top}) and the dust scattered halo
spectrum (\emph{left bottom}). Some residuals between the spectrum and
the model changing the gain and offset values are also shown (\emph{right bottom}).
}
\label{astroinstru1}
\end{figure*}

First of all, to discard the possibility that this absorption feature could be 
related to a gain effect, the data were fitted using different gain and offset 
values (up to $\pm$3\% and $\pm$0.05 keV, respectively).
The line profile expressed as the ratio of spectral data to a best-fit
model was always around 70\%, suggesting that it is not a gain effect
and that the absorption feature could be real. Figure~\ref{astroinstru1} (right bottom) shows
the residuals between the spectrum and the continuum model (three
absorbed power laws) in the energy range 0.9--8.0 keV.
 Then, in order to rule out the possibility of a calibration issue related
to the absorption feature,
a careful study of this observation as well as a sample of observations used
        by the \emph{XMM-Newton} cross calibration (\emph{XCAL}) archive was performed
        by the \emph{XCAL} team. The cross-calibration \emph{XMM-Newton}
database consists of $\sim$150 observations of different sources,
optimally reduced, fit with spectral models defined on a source-by-source
basis\footnote{http://www.iachec.org/meetings/2009/Guainazzi\_2.pdf}.
The conclusion of this calibration study was that the equivalent width of this
        feature (39$^{+11}_{-21}$ eV) is larger than the typical systematic uncertainties in this band (less than 5.3 eV) arguing against the
instrumental origin.

Then we look for possible astrophysical origins. We have extracted spectra from two blank regions (see Figure~\ref{astroinstru1} left top image) containing only
the background and the background plus halo, using the same extraction radius.
The spectrum of the dust scattered halo area has been corrected for the
background. As can be seen,
no absorption feature can be detected at 2.1 keV in any of them. The absorption feature appears
         only when the source is included in the extraction region suggesting a system origin. On the other hand,
we tested the possibility that the soft component was due to the presence
of an ionized absorber. Adopting the \textsc{Absori} model in \textsc{xspec}
(\cite{done92}), we obtained an unsatisfactory fit to the data and there
were still strong negative residuals around 2.1 keV.

In 4U 1538$-$52, assuming the observed feature is most likely intrinsic to the NS,
         there are two potential ways for the absorption feature to be generated in the NS
         atmosphere: cyclotron lines and atomic transition lines.
              \begin{itemize}
        \item \textbf{Cyclotron lines:} This hypothesis does not look plausible
because this system presents two electron-cyclotron lines at $\sim$21 keV and $\sim$47 keV (\cite{clark90,RR09}).
Alternatively, one can assume that this feature is associated with ion-cyclotron energies. However,
the surface magnetic field needed for this interpretation is greater than the magnetic field
inferred for the electron-cyclotron lines.
        \item \textbf{Atomic lines:} This possibility implies that the feature is formed
in the NS atmosphere. We can exclude
this feature as emerging from a hydrogen atmosphere because at any magnetic field and any
reasonable redshift, there is no pair of strong hydrogen spectral lines whose energies
would match the observed one. Therefore, one has to invoke heavier elements.
Another possibility is an iron atmosphere at B around 10$^{12}$ G
(\cite{mereghetti02}, \cite{moha06}), but the iron atmosphere should show many more
lines than the only one observed in the X-ray band (\cite{rajagopal}) and an
unreasonable value of gravitational redshift (\cite{moha06}).
An O/Ne atmosphere (He-like Oxigen, Li-like Oxygen) at B around 10$^{12}$ G is
plausible with the observed feature energy (\cite{moha06}). However, the O/Ne
atmosphere should show other absorption features at lower energies, but the
soft excess of the system prevents us any confirmation about their presence.
     \end{itemize}

\begin{figure}
\includegraphics[width=.45\textwidth]{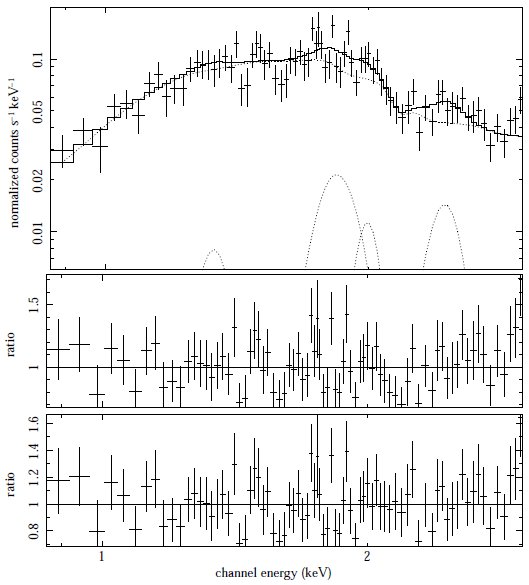}
\includegraphics[width=.5\textwidth]{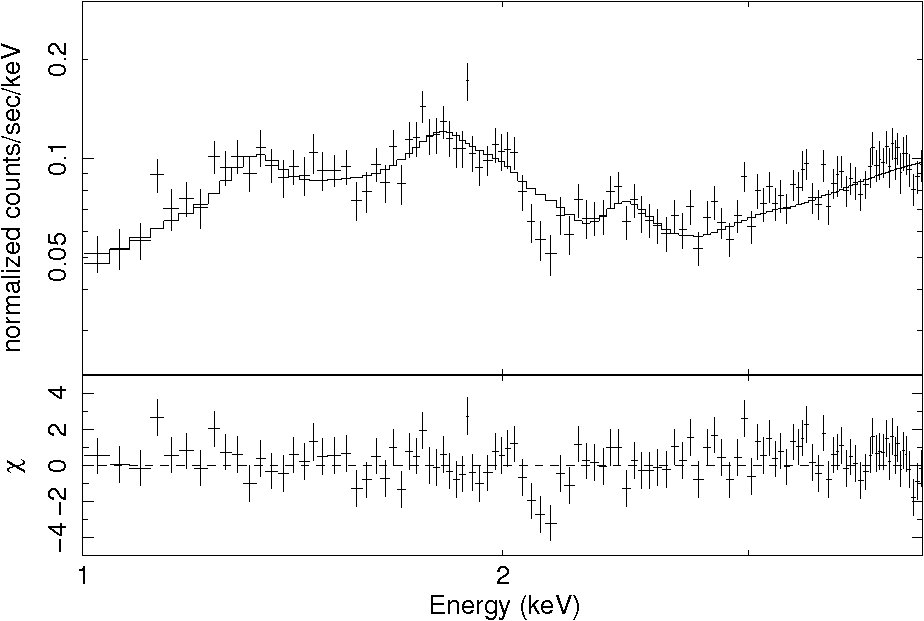}
\caption{Recombination emission lines of He- and H-like species
    in the eclipse spectrum of 4U 1538$-$52.
    \emph{Left top panel}: Spectrum and best fit model (an absorbed power-law modified
    by a Gaussian absorption line and
    four Gaussian emission lines) in the 0.8--3.0 keV energy range
    obtained with \emph{PN} camera.
    \emph{Left middle panel}: Residuals when comparing the data to a simple absorbed power-law
    plus four emission lines.
    \emph{Left bottom panel}: Residuals when the absorption line has been
    included in the model.
    \emph{Right panel:} Spectrum and best fit model (top) and residuals without
    the absorption line (bottom).}
\label{spectra}
\end{figure}

\section{Conclusions}

We have presented the spectral analysis of the HMXB 4U 1538$-$52 using
an \emph{XMM-Newton} observation focusing our attention specifically
within the energy range 0.8--3.0 keV. In this work, we show that the \emph{XMM} spectrum of 4U 1538$-$52 shows a deep significant
absorption feature at around 2.1 keV. Throughout a detailed analysis of the spectral
region covering the absorption line, we have been able to discard an instrumental origin.
Indeed, it is not present in any of the long term monitoring sources used for the \emph{XMM}
calibration. Likewise, it can not be accounted for by gain and/or offset corrections and the
latest calibration files.

Based on the spectral analysis we have derived the following conclusions:
     \begin{itemize}
        \item We conclude that the line is most likely of astrophysical origin and it is
formed in the wind or the atmosphere of the NS.
        \item An O/Ne atmosphere is plausible, although other absorption lines should
show at lower energies. The soft excess of the system and the \emph{RGS} low level of counts
prevents us any confirmation about their presence.
        \item Through a search in the main atomic lines databases we would have suggested
it is due to atomic transitions of hydrogen and helium like Fe or Si ions but,
existing models of ionized absorbers can not account for the line.
     \end{itemize}

\acknowledgments{This work was supported by the MICINN
  \emph{De INTEGRAL a IXO: binarias de rayos X y estrellas activas} project
  number AYA2010-15431. JJRR acknowledges the support by the
  Spanish Ministerio de Educaci\'on y Ciencia under grant
  PR2009-0455.}

\end{document}